\documentclass[aps,pra,10pt]{revtex4-2}

\usepackage{graphicx}
\usepackage{color}
\usepackage{amsmath}
\newcommand{\ket}[1]{\left| #1 \right\rangle}
\newcommand{\bra}[1]{\left\langle #1 \right|}
\newcommand{\proj}[1]{\ket{#1}\hskip-1mm\bra{#1}}

\newcommand{\braket}[2]{\langle #1|#2 \rangle}
\newcommand{\ketbra}[2]{\left|#1\right\rangle\hskip-1mm\left\langle#2\right|}

\begin{document}

\title{Statistical signatures of quantum contextuality}

\author{Holger F. Hofmann}
\email{hofmann@hiroshima-u.ac.jp}
\affiliation{
Graduate School of Advanced Science and Engineering, Hiroshima University,
Kagamiyama 1-3-1, Higashi Hiroshima 739-8530, Japan
}

\begin{abstract}
Quantum contextuality describes situations where the statistics observed in different measurement contexts cannot be explained by a measurement independent reality of the system. The most simple case is observed in a three-dimensional Hilbert space, with five different measurement contexts related to each other by shared measurement outcomes. The quantum formalism defines the relations between these contexts in terms of well-defined relations between operators, and these relations can be used to reconstruct an unknown quantum state from a finite set of measurement results. Here, I introduce a reconstruction method based on the relations between the five measurement contexts that can violate the bounds of non-contextual statistics. A complete description of an arbitrary quantum state requires only five of the eight elements of a Kirkwood-Dirac quasi probability, but only an overcomplete set of eleven elements provides an unbiased description of all five contexts. A set of five fundamental relations between the eleven elements reveals a deterministic structure that links the five contexts. As illustrated by a number of examples, these relations provide a consistent description of contextual realities for the measurement outcomes of all five contexts. 
\end{abstract}

\maketitle

\section{Introduction}

The interpretation of quantum mechanics is controversial because it is impossible to identify individual measurement outcomes with an underlying measurement independent reality \cite{Leg05,Han22}. Unfortunately, the familiar terminology of quantum mechanics tends to encourage images of underlying realities by carelessly identifying quantum state components with measurement outcomes even when no such measurements are ever performed. It cannot be stressed enough that quantum mechanics contradicts such fantasies directly by explaining the relation between different measurement contexts in terms of superpositions of their outcomes. Still, quantum mechanics does not seem to provide any good alternative. How can it be that we can describe the failure of objective realism in great detail, but cannot find any good alternative that would explain the relation between different measurement contexts in a manner consistent with the highly successful theoretical formalism? Perhaps we need a more appropriate formulation of the problem, so that we can identify the correct relations between different measurements without any speculation about unseen realities. 

A very good illustration of the problem was recently given by Frauchiger and Renner, who pointed out that the paradoxical correlations between quantum systems observed in a scenario similar to Hardy's version of a Bell non-locality prevent any consistent description of measurement processes within a quantum system \cite{Fra18,Har93}. The measurement problem can thus be identified with the problem of contextuality originally explored by Kochen and Specker \cite{Koc67}. To solve the measurement problem, we need to understand how the quantum formalism relates different contexts to each other. Fortunately, we can build on a vast literature concerning this problem. Here, we focus on a three dimensional Hilbert space, the most simple system in which Kochen-Specker contextuality can be demonstrated \cite{Cli93,Lei05,Kly08,Cab13}. The demonstration of contextuality requires a set of five different measurement contexts related to each other through shared measurement outcomes. The quantum formalism then predicts the violation of an inequality that would apply to all non-contextual theories of measurement. It is thus easy to prove that quantum theory must be contextual. However, surprisingly little effort has been made to relate the assumptions of non-contextual theories directly to the quantum formalism. Is it possible to formulate quantum theory in such a way that the difference between its predictions and non-contextual theories becomes more apparent? 

Conventional formulations of quantum mechanics tend to represent states in a specific context, effectively hiding the relations with other contexts in the coherences between orthogonal states that represent the different outcomes of a single measurement context. In fact, the origin of contextuality can always be traced to state independent relations between different contexts expressed in terms of fundamental relations between non-commuting operators \cite{Mer90,Yu15,Wae15,Pav23}. Nevertheless, the experimentally observable evidence can only be obtained in the form of measurement statistics associated with specific quantum states \cite{Amb13}. In the end, contextuality describes statistical relations between different measurements that apply to arbitrary quantum states. To better understand the origin of contextuality in the quantum formalism, it is therefore useful to investigate representations of quantum states that relate different measurement contexts to each other. This can be done by interpreting the quantum formalism as a generalized probability theory \cite{Spe08,Sha21,Sch24,Wag24a}. Not surprisingly, the results show that the quantum formalism introduces non-positive quasi probabilities into the statistics, which explains the differences between the predictions of quantum mechanics and the non-contextual hidden variable theories proposed as an alternative. The problem is that there is a wide range of possible quasi probabilities based on different selections of measurements or different operator expansions \cite{Hof20}. To identify an appropriate quasi probability, it is necessary to consider the specific relations between the different measurement contexts that characterize any given scenario. Here, I will take a look at the possible reconstruction of quantum states from experimentally accessible data, also known as quantum state tomography. Originally, these methods emphasized the similarity between quantum theory and classical statistics, e.g. when a semi classical analysis is used to reconstruct a phase space representation of the quantum state \cite{Smi93}. It has also been shown that weak measurements or a variety of joint measurements can be used to reconstruct a Kirkwood-Dirac distribution of two non-commuting observables \cite{Lun11,Lun12,Hof12,Hof14,The17,Wag24}. These approaches to quantum tomography show that the reconstruction of an unknown state can be achieved by applying methods of classical statistics to characterize the non-classical correlations between the outcomes of incompatible measurements. Non-positive quasi probabilities are obtained because these correlations cannot be explained by a positive valued probability distribution over the possible combination of incompatible outcomes. As observed very early in the history of quantum mechanics, any attempt to describe quantum states as joint probabilities of complementary observables necessarily results in such non-positive distributions \cite{Moyal,Dirac}.    

In the following, I introduce a method of tomography that relates five different contexts to each other, highlighting the statistical features responsible for the demonstration of quantum contextuality in a three level system \cite{Ji24}. The contexts involved in this tomographical reconstruction can be visualized using a recently introduced three-path interferometer \cite{Hof23}. Non-contextual realities correspond to hypothetical paths of a classical particle through the interferometer. The tomographic method provides the operator algebra needed to identify the different paths with elements of the context independent quantum statistics. Relations between different measurement contexts can then be identified with specific elements of the operator algebra. The results of the analysis show that the quantum state can be reconstructed from a set of only five Kirkwood-Dirac quasi probabilities. Each of these Kirkwood-Dirac terms describes the statistical weight of one of the eleven quasi realities describing the path of a classical particle. Operator algebra can be used to derive five relations that define the Kirkwood-Dirac terms of the remaining paths. These five relations between the different contexts express the deterministic structure of the Hilbert space formalism, preventing an arbitrary assignment of hidden realities even in cases where the individual measurement results would allow it. The framework established by the tomographic approach to quantum contextuality thus provides details of quantum contextuality that are hidden by the ``black box'' approach of no-go theorems.

Quantum theory makes precise statements about the relations between different contexts. These statements show that quantum mechanics is not constrained by the assumptions of a measurement independent reality. Instead, the uncertainties of the initial state obtain a different meaning in each of the different contexts. This is not just an interpretation of the formalism, the universality of tomographic relations indicate that these are fundamental laws of physics. By designing experiments with sufficient care, it is possible to characterize the specific dependence of reality on the type of measurement that will eventually be performed. The relations presented in the following are a first step towards a fully deterministic description of the physical world that resolves the measurement problem by identifying the precise role that the measurement context plays in shaping the observable reality of all objects. It should be noted that this kind of progress is made possible by the realization that there is no need for a measurement independent reality. The application of the formalism strongly suggests that the familiar structure of reality that we experience in our actual lives is sufficiently explained in terms of the deterministic relations between different measurements. This is a point that is often overlooked in discussions of the measurement problem such as \cite{Leg05,Han22}. We do not have any direct access to any microscopic realities, and our ability to perform measurements is limited by our macroscopic interventions in the world around us. If we cannot reconcile our actual experience with our descriptions of ``reality,'' we need to question the description and not the experience.

\section{Characterization of measurement contexts}
\begin{figure}[ht]
\begin{picture}(360,220)
\put(0,0){\makebox(360,220){\vspace{-3.5cm}
\scalebox{0.8}[0.8]{
\includegraphics{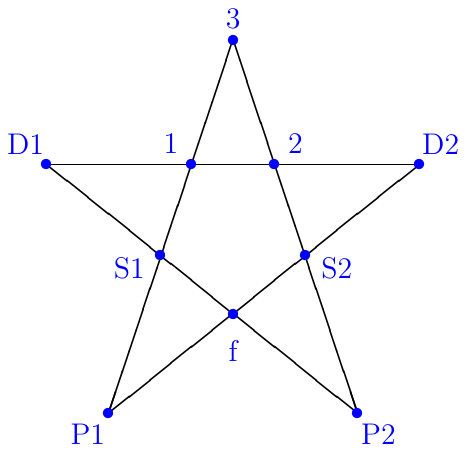}}}}
\end{picture}
\caption{\label{fig1}
Measurement contexts related by shared measurement results. The lines connect outcomes represented by orthogonal states, and each triangle represents a complete orthogonal basis. The cyclic relation between the five contexts results in inequalities that can be violated in quantum mechanics \cite{Kly08,Ji24}.
}
\end{figure}

Many fundamental investigations of quantum statistics start with the two dimensional Hilbert space of a logical qubit. The physical analogy is a spin-1/2 system, where every direction in space is a potential measurement basis. It is difficult to formulate a logical relation between the different measurement contexts, since no two measurements can be performed jointly and each measurement outcome belongs to only one possible measurement context. The situation changes when a three dimensional Hilbert space is considered. It is now possible to construct a much wider range of possible measurements, and two different measurement contexts can share the same measurement outcome. The Kochen-Specker theorem uses such relations to investigate the statistical relations between incompatible measurements, demonstrating that the relation between different measurement contexts cannot be described by uncertainty limited classical statistics \cite{Koc67}. In the most simple scenario, five different measurement contexts are related to each other by shared measurement outcomes in a three dimensional Hilbert space \cite{Cli93,Lei05,Kly08,Cab13}. Each measurement context is described by an orthogonal basis, where one of the basis states is unique to this context and the other two are shared with two other contexts, resulting in a cyclic relation between all five contexts. As shown in figure \ref{fig1}, the symmetry of this scenario can be illustrated by a pentagram, where each measurement context is represented by a triangle and the shared outcomes form a pentagon at the center. Note that this representation is a more complete version of the graph derived in \cite{Ji24}. I am using the same terminology, with the context $\{1,2,3\}$ at the top, the contexts $\{1,S1,D1\}$ and $\{2,S2,D2\}$ sharing the outcomes $1$ and $2$ at their respective sides, and the contexts $\{S1,f,P1\}$ and $\{S2,f,P2\}$ at the bottom, joined by the shared outcome $f$. A statstical proof of contextuality can now be formulated using only the five shared outcomes. Since each of these five outcomes excludes two of the others, non-contextual logic requires that no more than two of the shared outcomes can be true at the same time. The sum of the probabilities for these five outcomes should therefore have an upper bound of two. However, this restriction does not apply in Hilbert space, resulting in possible violations of the non-contextual bound.

\begin{figure}[ht]
\vspace{-1cm}
\begin{picture}(360,200)
\put(0,0){\makebox(360,200){\vspace{-3cm}
\scalebox{0.8}[0.8]{
\includegraphics{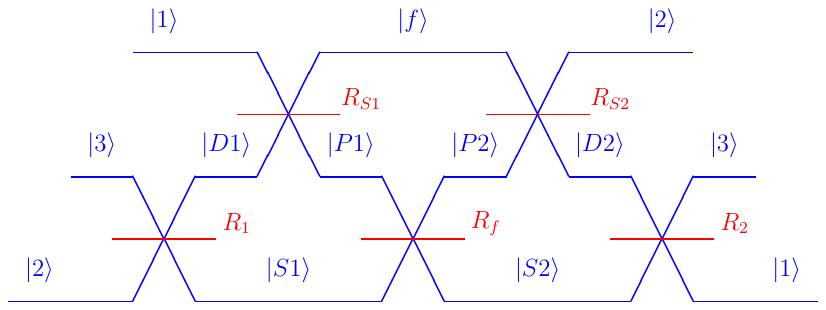}}}}
\end{picture}
\caption{\label{fig2}
Three-path interferometer illustrating the relation between the five different measurement contexts \cite{Hof23}. Non-contextual relations correspond to the paths of a classical particle through the interferometer. Beam splitter reflectivities represent Hilbert space inner products relating the coherent amplitudes of the output state basis to those of the input state basis. 
}
\end{figure}

The conventional formulation of quantum theory forces us to chose a basis in which to represent the states and the operators. This can cause serious misunderstandings, since each basis only represents one of the five possible measurements. It is important to remember that the physics of a quantum system should be independent of the basis chosen to describe it. Although figure \ref{fig1} correctly identifies the shared measurement outcomes, it remains unclear what the relations between the other outcomes are. In fact, quantum theory defines the relation between two measurement contexts in terms of a unitary transformation, and in the case of the five contexts shown in figure \ref{fig1}, the transformations between adjacent contexts involves a transformation in the two-dimenional subspace orthogonal to the shared context \cite{Ji24}. As I have shown recently, this relation can be illustrated by a three-path interferomer, where the measurement outcomes of the different contexts are represented by the paths along which a single quantum particle propagates through the interferometer \cite{Hof23}. The transformation between adjacent contexts is then implemented by beam splitters that control the overlap between the quantum states representing the paths. Figure \ref{fig2} shows the interferometer. Each of the five beam splitters can be identified with the path parallel to it, corresponding to the shared measurement outcome of the contexts before and after the beam splitter. As was shown in \cite{Hof23,Ji24}, selecting any two beam splitter reflectivities will determine the other three. An option that is particularly easy to analyze is obtained when $R_1=R_2=1/2$, which results in the remaining reflectities being given by $R_{S1}=R_{S2}=1/3$ and $R_f=1/4$.

\begin{table}[ht]
 
    \label{tab1}
    \centering \large
    \begin{tabular}{|cc||ccc|ccc|ccc|ccc|ccc||}
    \multicolumn{17}{c}{}
    \\ \hline
       Path  &&&\multicolumn{13}{c}{State Vectors}& \\ \hline \hline
        $[0]$&&& $\ket{3}$ &&& $\ket{D1}$ &&& $\ket{P1}$ &&& $\ket{P2}$ &&&$\ket{D2}$ & \\ \hline \hline
        $[1,P2]$&&& \multicolumn{5}{c|}{$\ket{1}$}&&$\ket{P1}$&&& $\ket{P2}$ &&& $\ket{D2}$ & \\ \hline
        $[S1,D2]$&&& $\ket{3}$ &&& \multicolumn{4}{c}{$\ket{S1}$} &&& $\ket{P2}$&&& $\ket{D2}$ & \\ \hline
        $[f,3]$&&& $\ket{3}$ &&& $\ket{D1}$ &&&\multicolumn{4}{c}{$\ket{f}$}&&&$\ket{D2}$&  \\ \hline
        $[S2,D1]$&&& $\ket{3}$&&& $\ket{D1}$ &&& $\ket{P1}$ &&& \multicolumn{4}{c}{$\ket{S2}$} & \\ \hline
        $[2,P1]$&&& $\ket{2}$ &&& $\ket{D1}$ &&& $\ket{P1}$ &&& $\ket{P2}$ &&& $\ket{2}$&  \\ \hline \hline
        $[1,f]$&&& \multicolumn{4}{c}{$\ket{1}$}&&&\multicolumn{4}{c}{$\ket{f}$}&&& $\ket{D2}$ & \\ \hline
$[S1,S2]$&&& $\ket{3}$ &&& \multicolumn{4}{c}{$\ket{S1}$}&&&\multicolumn{4}{c}{$\ket{S2}$} & \\ \hline
        $[2,f]$&&& $\ket{2}$ &&& $\ket{D1}$ &&&  \multicolumn{4}{c}{$\ket{f}$}&&& $\ket{2}$ & \\ \hline
$[1,S2]$&&& \multicolumn{4}{c}{$\ket{1}$}&&& $\ket{P1}$ &&& \multicolumn{4}{c}{$\ket{S2}$} & \\ \hline
  $[2,S1]$&&& $\ket{2}$ &&& \multicolumn{4}{c}{$\ket{S1}$}&&&  $\ket{P2}$ &&& $\ket{2}$ & \\ \hline \hline
    \end{tabular}
    \caption{State vectors associated with the eleven paths through the three-path interferometer, corresponding to the notion of a non-contextual reality of all five contexts. The states belonging to one path all have non-zero inner products. One of the paths has five segments, five of the paths have four segments, and five of the paths have three segments.}
\end{table}

What can be said about the relation between different measurement contexts? The identification of measurement contexts with the different stages of single-particle propagation in a three-path interferometer illustrates this relation by associating measurement outcomes with physical paths. A non-contextual hidden variable model would then assign a specific path through the interferometer to each particle. Since the input context $\{1,2,3\}$ is perfectly recovered at the output ports, paths originating from input $1$ must end up in output $1$ and so forth, leaving a total of eleven distinct possibilities for the propagation of the particle. Table \ref{tab1} illustrates the eleven paths by assigning a sequence of Hilbert space vectors to each of them. As the table shows, two path segments $[a,b]$ are sufficient to define each of the paths, except for the path composed of the five segments that are unique to each context, which is therefore labeled $[0]$. The quantum mechanical relations that replace the sequence of paths in the Hilbert space formalism are inner products of the state vectors that represent the measurement outcomes of the different contexts \cite{Ji24}. Although quantum contextuality shows that quantum states cannot be represented by positive valued probability distributions over the eleven paths in table \ref{tab1}, it should be possible to find a quasi probability representation of the density matrix that relates directly to these $11$ paths. It has already been noted elsewhere that Kirkwood-Dirac quasi probabilities are ideally suited for this purpose \cite{Hof23,Han24,Ji24b}. In the following, I will develop an expansion of the density matrix that can serve as a starting point for a detailed investigation of the relations between the five measurement contexts described by the Hilbert space formalism.

\section{Relations between non-orthogonal states}

As explained in previous work, Kirkwood-Dirac distributions are the most consistent representation of quantum statistics, accurately reproducing the relations between non-orthogonal states in the Hilbert space formalism \cite{Hof11,Hof12b,Hof14b,Hof15,Hal18,Bud23,Wag23,Los23,Ume24}. Such relations represent the non-classical structure of operator statistics and ensure that quantum theory cannot be reconciled with any classical statistical theory \cite{Hof20,Hof15}.  Here, the goal is to identify a specific set of relations between five different measurement contexts that make optimal use of the orthogonality relations used to demonstarte quantum contextuality. To achieve this, we can start with a selection of non-orthogonal states representing one of the five trajectories composed of three states, e.g. $\{\ket{1},\ket{f},\ket{D2}\}$. These three state vectors form a complete non-orthogonal basis of the three-dimensional Hilbert space. The corresponding set of contravariant vectors is $\{\ket{S2},\ket{2},\ket{S1}\}$, and the expansion of any state vector in this non-orthogonal basis can be expressed by representing the identity operator as a sum of three non-hermitian operators,
\begin{equation}
\label{eq:identity}
\frac{\ketbra{1}{S2}}{\braket{S2}{1}} + \frac{\ketbra{f}{2}}{\braket{2}{f}} + \frac{\ketbra{D2}{S1}}{\braket{S1}{D2}} = \hat{\rm{I}}.
\end{equation}
The decomposition of the identity operator $\hat{\rm{I}}$ into non-hermitian operators provides us with an interesting alternative to the decomposition into projectors of orthogonal states used in conventional expansions of the quantum state. The individual non-hermitian operators in Eq.(\ref{eq:identity}) can be used to represent the pre- and post-selected ensembles of weak measurements \cite{Hof10} and appear in the reconstruction of quantum states using direct measurements of the Kirkwood-Dirac distribution \cite{Lun11,Lun12,Hof12b}. When interpreted statistically, these operators thus represent the quantum mechanical analog of a logical AND. To simplify the terminology, we can define the operator
\begin{equation}
\hat{\Lambda}(a,b):= \frac{\ketbra{a}{b}}{\braket{b}{a}}.
\end{equation}
As shown in \cite{Hof10}, these operators describe the weak measurement statistics of an initial state $\ket{a}$ under the post-selection condition $\ket{b}$. The expectation values of these operators for the states $\ket{a}$ and $\ket{b}$ are necessarily one. Likewise, they are necessarily zero for every state that is orthogonal to either $\ket{a}$ or $\ket{b}$. These properties explain why the identity operator can be decomposed in the manner shown in Eq.(\ref{eq:identity}). In total, there are five operator relations of this kind, 
\begin{eqnarray}
\label{eq:Lambdas}
\hat{\Lambda}(1,S2) + \hat{\Lambda}(f,2) + \hat{\Lambda}(D2,S1) &=& \hat{\rm{I}}
\nonumber \\ 
\hat{\Lambda}(S1,2) + \hat{\Lambda}(S2,1) + \; \hat{\Lambda}(3,f) \;\; &=& \hat{\rm{I}}
\nonumber \\
\hat{\Lambda}(f,1) + \hat{\Lambda}(2,S1) + \hat{\Lambda}(D1,S2) &=& \hat{\rm{I}}
\nonumber \\
\hat{\Lambda}(S2,S1) + \hat{\Lambda}(1,f) + \hat{\Lambda}(P1,2) &=& \hat{\rm{I}}
\nonumber \\
\hat{\Lambda}(2,f) + \hat{\Lambda}(S1,S2) + \hat{\Lambda}(P2,1) &=& \hat{\rm{I}}
\end{eqnarray}
It should be noted that each combination $(a,b)$ corresponds to one of the paths in table \ref{tab1}. The three operators in each relation represent paths that do not share the same outcome in any of the five contexts. When the basis states of one of the contexts are used to define the matrix representations of the operators, the diagonal elements will be one for the state that is part of the path, and zero for the two other states. For example, the state $\ket{3}$ is part of paths $[S1,D2]$, $[f,3]$, $[S2,D1]$ and $[S1,S2]$. Therefore,
\begin{equation}
    \bra{3}\hat{\Lambda}(S1,D2)\ket{3}=\bra{3}\hat{\Lambda}(f,3)\ket{3}=\bra{3}\hat{\Lambda}(D2,S1)\ket{3}=\bra{3}\hat{\Lambda}(S1,S2)\ket{3}=1.
\end{equation}
For all other operators $\hat{\Lambda}(a,b)$,
\begin{equation}
    \bra{3}\hat{\Lambda}(1,P2)\ket{3}=\bra{3}\hat{\Lambda}(2,P1)\ket{3}=\bra{3}\hat{\Lambda}(1,f)\ket{3}=\bra{3}\hat{\Lambda}(2,f)\ket{3}=\bra{3}\hat{\Lambda}(1,S2)\ket{3}=\bra{3}\hat{\Lambda}(2,S1)\ket{3}=0.
\end{equation}
The operators $\hat{\Lambda}(a,b)$ thus provide a quantum mechanical representation of $10$ of the $11$ paths shown in table \ref{tab1}. In the following, these operators and the related expansion of arbitrary quantum states will be used to derive a method of quantum tomography that highlights the symmetry between the five different contexts using a multi-context version of a Kirkwood-Dirac quasi probability.

\section{Complete expansions of arbitrary states}

The completeness relation in Eq.(\ref{eq:identity}) can be used to express any pure state $\ket{\psi}$ as a superposition of the three states $\{\ket{S2},\ket{2},\ket{S1}\}$, where the coefficients are given by inner products with the states $\{\ket{1},\ket{f},\ket{D2}\}$,
\begin{equation}
\label{eq:complete}
    \ket{\psi}= \frac{\braket{1}{\psi}}{\braket{1}{S2}} \ket{S2} + \frac{\braket{f}{\psi}}{\braket{f}{2}} \ket{2} + \frac{\braket{D2}{\psi}}{\braket{D2}{S1}} \ket{S1}. 
\end{equation}
It should be noted that non-orthogonal expansions reveal a fundamental problem with the notion that each of the components represents a possible ``reality''. The coefficients of the expansion are not zero when the state is orthogonal to that component, but instead refer to a different set of states. This means that expansions of states orthogonal to one or two of the components will include non-zero contributions of those components. The problem is best illustrated by the expansion of the state $\ket{D2}$,
\begin{equation}
\label{eq:D2expand}
    \ket{D2} = \frac{\braket{1}{D2}}{\braket{1}{S2}} \ket{S2} + \frac{\braket{f}{D2}}{\braket{f}{2}} \ket{2} + \frac{1}{\braket{D2}{S1}} \ket{S1}. 
\end{equation}
Destructive interference between $\ket{2}$ and $\ket{S1}$ ensures that $\braket{S2}{D2}=0$, and destructive interference between $\ket{S2}$ and $\ket{S1}$ ensures that $\braket{2}{D2}=0$. This requires non-trivial relations between the inner products,
\begin{eqnarray}
\braket{1}{D2} \braket{D2}{S1} &=& - \braket{1}{S2} \braket{S2}{S1} 
\nonumber \\
 \braket{f}{D2} \braket{D2}{S1} &=& - \braket{f}{2} \braket{2}{S1}    
\end{eqnarray}
As this example shows, all of the inner products are related to each other by the orthogonality relations between the five different measurement contexts \cite{Ji24}. To simplify the relations, it is useful to introduce a specific numerical example. As noted in \cite{Ji24}, the example can be defined by the expansion of the states $\ket{S1}$ and $\ket{S2}$ in the $\{\ket{1},\ket{2},\ket{3}\}$ basis. It is therefore convenient to chose equal superpositions,
\begin{eqnarray}
\label{eq:numerics}
   \ket{S1}&=&\frac{1}{\sqrt{2}} \left(\ket{2} + \ket{3}\right)
\nonumber \\
   \ket{S2}&=&\frac{1}{\sqrt{2}} \left(\ket{1} + \ket{3}\right).
\end{eqnarray}
All other superpositions can then be derived using the orthogonality conditions of the relations between the five measurement contexts. Most notably, $\ket{f}$ is orthogonal to both $\ket{S1}$ and $\ket{S2}$, so the representation of the state in the $\{\ket{1},\ket{2},\ket{3}\}$ basis is necessarily given by
\begin{equation}
    \ket{f} = \frac{1}{\sqrt{3}} \left(\ket{1}+\ket{2}-\ket{3} \right).
\end{equation}
With these definitions, the expansion of $\ket{D2}$ in Eq.(\ref{eq:D2expand}) is given by
\begin{equation}
\label{eq:illustrate}
    \ket{D2} = \ket{S2} + \sqrt{2} \ket{2} - 2 \ket{S1}. 
\end{equation}
Note that the physics represented by the state $\ket{D2}$ cannot depend on its representation. The characterization of an unknown state should not be based on the components of a representation, but on the coefficients of the expansion and any physical meaning that they may have that is independent of the expansion in which they are used.

The coefficients of the non-orthogonal expansion of quantum states given by Eq.(\ref{eq:complete}) describe coherences between non-orthogonal states. Such coherences can be represented by Kirkwood-Dirac quasi probabilities,
\begin{equation}
\label{eq:KDterm}
    \varrho(a,b)=\braket{b}{a}\bra{a}\hat{\rho}\ket{b}.
\end{equation}
As mentioned before, the Kirkwood-Dirac quasi probabilities describe the relation between the quantum statistics of $\ket{a}$ and $\ket{b}$ in terms of the weak values of $\proj{a}$ post-selected in $\ket{b}$ \cite{Lun11,Lun12,Dia15}.  In the following, I will refer to them as KD terms. When the non-orthogonal expansion of Eq.(\ref{eq:complete}) is applied to a density operator $\hat{\rho}$, the coefficients of the expansion can be expressed by a combination of probabilities and KD terms, 
\begin{eqnarray}
\label{eq:expand}
\hat{\rho} &=& 2 P(1) \proj{S2} + 3 P(f) \proj{2} + 4 P(D2) \proj{S1} 
+ 3 \sqrt{2} \varrho(1,f) \ketbra{S2}{2} + 3 \sqrt{2} \varrho(f,1) \ketbra{2}{S2}
\nonumber \\[0.1cm] && 
- 2 \varrho(1,D2) \; \hat{\Lambda}(S2,S1) - 2 \varrho(D2,1) \; \hat{\Lambda}(S1,S2)
- 3 \varrho(f,D2) \; \hat{\Lambda}(2,S1) - 3 \varrho(D2,f) \; \hat{\Lambda}(S1,2).
\end{eqnarray}
The expansion combines the probabilities of the three outcomes $1$, $f$ and $D2$ with KD terms for all three combinations, $(1,f)$, $(1,D2)$ and $(f,D2)$. These six elements correspond to the diagonal and off-diagonal terms of a density matrix, where the coherences are replaced by complex quasi probabilities that assign a joint statistical weight to each pair of outcomes. 

The expansion given in Eq.(\ref{eq:expand}) did not require any normalization of the density operator $\hat{\rho}$. It is therefore possible to derive an additional relation between the six coefficients using the condition $\mbox{Tr}(\hat{\rho})=1$. For example, it is possible to express the probability of $D2$ as
\begin{equation}
\label{eq:D2result}
P(D2) = \frac{1}{4}\left(1-2 P(1) - 3 P(f) + \mbox{Re}\left(4 \varrho(1,D2) + 6 \varrho(D2,f)\right)\right).    
\end{equation}
A normalized quantum state is completely determined by the two probabilites $P(1)$, $P(f)$, and the three KD terms $\varrho(1,f)$, $\varrho(1,D2)$, $\varrho(D2,f)$. Experimentally, KD terms can be determined by weak measurements or by a variety of other methods involving quantum interferences \cite{Lun11,Lun12,Hof12,Hof14,The17,Wag24}. Eqs.(\ref{eq:expand}) and (\ref{eq:D2result}) show that the statistics of all possible measurements can be expressed by linear combinations of these five coefficients and the normalization. In the following, we will take a look at how this expansion of the quantum state relates different contexts to each other.

\section{Reconstruction of a Kirkwood-Dirac distribution}
\label{sec:reconstruct}

A Kirkwood-Dirac distribution describes the relation between two different measurement contexts in the form of a complex joint probability of all possible combinations of outcomes. If the inner products of outcomes from different contexts are non-zero, the Kirkwood-Dirac distribution of these contexts is a complete representation of the quantum state \cite{Lun11,Lun12,Hof11,Hof12b}. However, this procedure cannot be applied to any of the five contexts, since at least two of the outcomes from different contexts will be orthogonal to each other. It is therefore not surprising that the expansion in Eq.(\ref{eq:expand}) combines outcomes from all five contexts. The selection of the specific expansion given by Eq.(\ref{eq:identity}) defines expansion coefficients associated with the path $[1,f]$ in table \ref{tab1}. It is possible to relate these coefficients to a complete Kirkwood-Dirac distribution for the contexts defined by the unique outcomes $\ket{3}$ and $\ket{P2}$. This Kirkwood-Dirac distribution has a total of eight non-zero elements, as shown in table \ref{tab2}. Five of these elements are needed to determine the five coefficients of the expansion in Eq.(\ref{eq:expand}). It is conventient to use the five coefficients marked in red, since the coefficients can then be determined by simple sums. The probabilities $P(1)$ and $P(f)$ are marginals of the distribution,
\begin{eqnarray}
\label{eq:marginals}
    P(1) &=& \varrho(1,f) + \varrho(1,S2) + \varrho(1,P2),
    \nonumber \\
    P(f) &=& \varrho(1,f) + \varrho(2,f) + \varrho(3,f).
\end{eqnarray}
The remaining two coefficients are given by KD terms that involve $D2$. Since $\ket{S2}$ is orthogonal to $\ket{D2}$, $\varrho(1,D2)$ does not include the contribution of $\varrho(1,S2)$ to $P(1)$. The same holds true for $\varrho(D2,f)$ and $\varrho(2,f)$. The two remaining KD terms in the expansion are therefore given by
\begin{eqnarray}
   \varrho(1,D2) &=& \varrho(1,f) + \varrho(1,P2), 
\nonumber \\
   \varrho(D2,f) &=& \varrho(1,f) + \varrho(3,f).
\end{eqnarray}
Only five of the eight elements of the Kirkwood-Dirac distribution for the contexts $\{\ket{1}, \ket{2}, \ket{3}\}$ and $\{\ket{f}, \ket{S2}, \ket{P2}\}$  are needed to completely define an arbitrary quantum state. The relation between the contexts represented by the orthogonality of $\ket{2}$ and $\ket{S2}$ thus simplifies the characterization of an unknown state. 

\begin{table}[ht]
    \centering \large
    \begin{tabular}{|cc||ccc|ccc|ccc||}
    \multicolumn{11}{c}{}
    \\ \hline
            
        &&& $\ket{1}$ &&& $\ket{2}$ &&& $\ket{3}$ & \\ \hline \hline
        $\ket{f}$&&& {\color{red} $\varrho(1,f)$} &&& {\color{red}$\varrho(2,f)$} &&& {\color{red}$\varrho(3,f)$} & \\ \hline
        $\ket{S2}$&&& {\color{red}$\varrho(1,S2)$} &&& {\bf -----} &&& $\varrho(3,S2)$ & \\ \hline
        $\ket{P2}$&&& {\color{red}$\varrho(1,P2)$} &&& $\varrho(2,P2)$ &&& $\varrho(3,P2)$ & 
        \\ \hline \hline
    \end{tabular}
    \caption{Kirkwood-Dirac distribution of the contexts $\ket{a}=\{\ket{1},\ket{2},\ket{3}\}$ and $\ket{b}=\{\ket{f},\ket{S2},\ket{P2}\}$. The five KD terms highlighted in red are sufficient for a complete reconstruction of an unknown state.}
    \label{tab2}
\end{table}

The Kirkwood-Dirac distribution is a helpful representation of the relations between two different contexts. It is therefore interesting to investigate how Eq.(\ref{eq:expand}) relates the five coefficients that appear in the expansion to the remaining three elements of the Kirkwood-Dirac distribution in table \ref{tab2}. The easiest way to find the result is to consider the probabilities of $S2$ and $2$,
\begin{eqnarray}
\label{eq:P2PS2}
    P(2) &=& \frac{1}{2}\left(1-2 P(1) + 3 P(f) + \mbox{Re}\left(4 \varrho(1,D2)-6 \varrho(D2,f)\right)\right)
    \nonumber \\
    P(S2) &=& \frac{1}{4}\left(1+6 P(1) - 3 P(f) - \mbox{Re}\left(12 \varrho(1,D2)-6 \varrho(D2,f)\right)\right). 
\end{eqnarray}
These marginals of the Kirkwood-Dirac distribution determine the KD terms $\varrho(3,S2)$ and $\varrho(2,P2)$ from the elements marked in red in table \ref{tab2}. The remaining element $\varrho(3,P2)$ can be determined from the normalization of the distribution. Since $\ket{2}$, $\ket{S2}$ and $\ket{D2}$ form the orthogonal basis of a third context, it is also possible to determine $\varrho(3,P2)$ from the probability of $D2$ as given in Eq.(\ref{eq:D2result}). The three probabilities are related to the elements of the Kirkwood-Dirac distribution by
\begin{eqnarray}
P(2) &=& \varrho(2,f) + \varrho(2,P2),
\nonumber \\
P(S2) &=& \varrho(1,S2) + \varrho(3,S2),
\nonumber \\
P(D2) &=& \varrho(1,f) + \varrho(3,f) + \varrho(1,P2) + \varrho(3,P2).
\end{eqnarray}
The remaining elements of the Kirkwood-Dirac distribution can then be expressed by the five KD terms used in the reconstruction of the quantum state,
\begin{eqnarray}
\label{eq:KDS1}
    \varrho(3,P2) &=& \frac{1}{4}\left(1+\varrho(1,f)-3\varrho(2,f)-\varrho(3,f)- 2\varrho(1,S2)-2 \varrho(1,P2)\right),
    \nonumber \\
    \varrho(2,P2) &=& \frac{1}{2}\left(1- \varrho(1,f)+ \varrho(2,f)- 3 \varrho(3,f) - 2 \varrho(1,S2)+ 2 \varrho(1,P2)\right),
    \nonumber \\
    \varrho(3,S2) &=& \frac{1}{4}\left(1- 3 \varrho(1,f) - 3 \varrho(2,f) + 3 \varrho(3,f) +2 \varrho(1,S2) - 6 \varrho(1,P2)\right).
\end{eqnarray}
These equations represent a set of universal rules that relate the elements of a single Kirkwood-Dirac distribution to each other. It is worth looking at these rules in detail, to figure out what kind of statistical relations they express.

The most obvious reason why there should be a relation between the elements is the normalization of the distribution. Probabilities attributed to the five elements used in the reconstruction will be missing from the three additional elements. If all of the elements used in the reconstruction are zero, the state is $\ket{S1}$ and the three additional elements are $\varrho(3,P2)=1/4$, $\varrho(2,P2)=1/2$, and $\varrho(3,S2)=1/4$. It is interesting to observe that there is no degree of freedom left. There is only one state with $P(1)=P(f)=0$. Eq.(\ref{eq:KDS1}) then defines the necessary uncertainty of this state in terms of a joint probability of the remaining outcomes $(2,3)$ and $(S2,P2)$. We can also consider cases where there is only one non-zero element among the five elements used in the reconstruction of the state. Most obviously, this is the case for $\ket{2}$, where $\varrho(2,f)=1/3$, and for $\ket{S2}$, where $\varrho(1,S2)=1/2$. In both cases, a contribution proportional to the non-zero element is subtracted from two of the three elements in Eq.(\ref{eq:KDS1}), leaving only one additional element with a non-zero value. Naturally, this element obtains exactly the value needed for a total probability of one, $\varrho(2,P2)=2/3$ for $\ket{2}$ and $\varrho(3,S2)=1/2$ for $\ket{S2}$. It is worth noting that a value of $\varrho(2,f)$ greater than $1/3$ would result in a negative value of $P(S2)$, indicating a strict limit to the concentration of quasi probabilities in a single KD term. A lower value of $\varrho(2,f)$ is possible, since it corresponds to a mixed state of $\ket{S1}$ and $\ket{2}$. The reason for the upper limit of $\varrho(2,f)$ is that the reconstructed state would be a non-positive state - specifically, a mixture of $\ket{2}$ and $\ket{S1}$ with a negative contribution from $\ket{S1}$. 

Finally, it may be instructive to consider states with particularly high probabilities of $1$ and $f$, since the reconstruction method and the Kirkwood-Dirac distribution put special emphasis on these two outcomes. For $\ket{f}$, the KD-terms are $\varrho(1,f)=1/3$, $\varrho(2,f)=1/3$ and $\varrho(3,f)=1/3$, while $\varrho(1,S2)=\varrho(1,P2)=0$. As expected, Eq.(\ref{eq:KDS1}) then states that the remaining three elements of the distribution are zero. Any attempt to re-distribute probabilities between the three non-zero elements will result in a re-distribution of a total probability of zero between $P(S2)$ and $P(P2)$, resulting in at least one negative probability. Eq.(\ref{eq:KDS1}) thus allows only one probability distribution of $1$, $2$ and $3$ for $P(f)=1$. The same holds true for the state $\ket{1}$, where the only possible probability distribution is $\varrho(1,f)=1/3$, $\varrho(1,S2)=1/2$ and $\varrho(1,P2)=1/6$, while $\varrho(2,f)=\varrho(3,f)=0$. 

Since we are interested in the relations between the two contexts in the Kirkwood-Dirac distribution, we can now take a look at a state that achieves high probabilities for both $1$ and $f$ simultaneously. Since such a state closely approximates the trajectory $[1,f]$, I will refer to it as the state $\ket{T_{1f}}$. To keep the numerics simple, let us consider the superposition
\begin{eqnarray}
    \ket{T_{1f}} &:=& \sqrt{\frac{4}{11}} \ket{1} + \sqrt{\frac{3}{11}} \ket{f} \nonumber \\ &=& \frac{1}{\sqrt{11}}\left(3\ket{1}+\ket{2}-\ket{3}\right).
\end{eqnarray}
This state has $P(S1)=0$, so that it can be represented as a superposition of $\ket{1}$ and $\ket{f}$. The interference between these two components is constructive, adding a probability of $4/11$ to the probabilities associated with the expansion coefficients. The state can also be written as
\begin{equation}
    \proj{T_{1f}} = \frac{4}{11} \proj{1} + \frac{3}{11} \proj{f} + \frac{2}{11} \hat{\Lambda}(1,f) + \frac{2}{11} \hat{\Lambda}(f,1). 
\end{equation}
When written in this form, it is easy to derive the probabilities for any of the five measurement contexts. For $\ket{1}$, we get contributions of $1$ from all operators except for $\proj{f}$, where the contribution is $1/3$. The total probability is therefore $P(1)= 9/11$. Similarly, $\ket{f}$ results in contributions of $1$ except for $\proj{1}$, where the contribution is 1/3. The total probability is $P(f)=25/33$. In classical statistics, these probabilities would suggest a joint probability of $1$ and $f$ of at least $19/33$, obtained under the assumption that the system is always either in $f$ or in $1$. However, the corresponing KD term is only $\varrho(1,f)=15/33$. This is possible because negative values appear in the KD terms that are neither related to $1$ nor to $f$.

\begin{table}[ht]
    \centering \large
    \begin{tabular}{|cc||ccc|ccc|ccc||}
    \multicolumn{11}{c}{}
    \\ \hline
            
        &&& $\ket{1}$ &&& $\ket{2}$ &&& $\ket{3}$ & \\ \hline \hline
        $\ket{f}$&&& 15/33 &&& 5/33 &&& 5/33 & \\ \hline
        $\ket{S2}$&&& 9/33 &&& {\bf -----} &&& -3/33 & \\ \hline
        $\ket{P2}$&&& 3/33 &&& -2/33 &&& 1/33 & 
        \\ \hline \hline
    \end{tabular}
    \caption{Kirkwood-Dirac distribution of the state $\ket{T_{1f}}$. Quantum coherences between $\ket{1}$ and $\ket{f}$ increase $\varrho(1,f)$ and reduce $\varrho(S2,3)$ and $\varrho(2,P2)$ to negative values.}
    \label{tab3}
\end{table}

Table \ref{tab3} shows the complete Kirkwood-Dirac distribution of the state $\ket{T_{1f}}$. According to Eq.(\ref{eq:KDS1}), an increase in $\varrho(1,f)$ will result in a decrease of $\varrho(3,S2)$ and $\varrho(2,P2)$. In the present case, we find that $\varrho(3,S2)=-3/33$ and $\varrho(2,P2)=-2/33$. The third KD term in Eq.(\ref{eq:KDS1}) has a positive value of $\varrho(3,P2)=1/33$. Eq.(\ref{eq:KDS1}) shows that these three values correspond an addition of $+4/33$ to $\varrho(1,f)$ compared with a situation where all three KD terms are initially zero, the situation obtained for any incoherent mixture of $\ket{1}$ and $\ket{f}$. The superposition state $\ket{T_{1f}}$ thus corresponds to a mixture of $\ket{1}$ and $\ket{f}$ with an addition of $+4/33$ to $\varrho(1,f)$. Specifically, a mixture of $6/11$ times $\proj{1}$ and $5/11$ times $\proj{f}$ has the same values for $\varrho(1,S2)$, $\varrho(1,P2)$, $\varrho(2,f)$ and $\varrho(3,f)$ as $\ket{T_{1f}}$, but all of the remaining probability are found in $\varrho(1,f)=11/33$. The coherence between $\ket{1}$ and $\ket{f}$ increases the value of $\varrho(1,f)$ by $+4/33$, and this increase in $\varrho(1,f)$ results in increase of $1/4$ times $4/33$ or $1/33$ in $\varrho(3,P2)$ a decrease of $-1/2$ times $4/33$ or $-2/33$ in $\varrho(2,P2)$, and a decrease of $-3/4$ times $4/33$ or $-3/33$ in $\varrho(3,S2)$. Eq.(\ref{eq:KDS1}) thus shows that negative KD terms are needed to achieve high values in other KD terms. The analysis in this section has shown that these relations are already obtained when only two contexts are related to each other. This is especially interesting because the Kochen-Specker theorem assigns a cyclic and symmetric role to all five contexts, as illustrated in figure \ref{fig1}. In the following, I will take a look at the inequality violation used to demonstrate contextuality in the present scenario and use it as a starting point for a formulation of cyclic relations between KD terms that correspond to non-contextual hidden variable theories for the five contexts.

\section{Inequality violations and cyclic relations between all five contexts}

Non-contextual hidden variable models assume that the outcomes of all measurements are pre-determined for each individual system. For the three-path interferometer in figure \ref{fig2}, this means that each particle follows a specific path through the five contexts, corresponding to the paths shown in table \ref{tab1}. If this is indeed the case, it is possible to identify extremal values of collective properties that define inequalities for the experimentally observable statistics. In the well-studied case of the five contexts defined here, the maximal number of shared outcomes that a path may include is two. This means that two is the maximal value we should obtain if we sum the probabilities of all shared outcomes \cite{Kly08},
\begin{equation}
\label{eq:ineqality}
    P(1)+P(2)+P(S1)+P(S2)+P(f) \leq 2.
\end{equation}
It is well known that this inequality can be violated, and various formulations of this paradox have been discussed \cite{Kly08,Cab13}. Here, our goal is to relate the paradox to the KD terms that describe the relations between different measurement contexts. 

In the tomographic reconstruction considered above, we have direct access to $P(1)$ and $P(f)$, as shown in Eq.(\ref{eq:marginals}). $P(2)$ and $P(S2)$ are given in Eq.(\ref{eq:P2PS2}). Since only the sum of $P(2)$ and $P(S2)$ is needed, it is sufficient to refer to $P(D2)$ in Eq.(\ref{eq:D2result}) to obtain
\begin{equation}
    P(2)+P(S2) = \frac{3}{4} + \frac{1}{2} P(1) + \frac{3}{4} P(f) - \mbox{Re}\left(\varrho(1,D2)+\frac{3}{2} \varrho(D2,f) \right).
\end{equation}
This leaves only the probability $P(S1)$. Using Eq.(\ref{eq:expand})and Eq.(\ref{eq:D2result}), this probability can be given as
\begin{equation}
    P(S1) = 1 - \frac{3}{2} P(1) - \frac{3}{2} P(f) + 3 \mbox{Re}(\varrho(1,f)).
\end{equation}
The sum of all five probabilities is then given by
\begin{eqnarray}
    \Sigma &:=&  P(1)+P(2)+P(S1)+P(S2)+P(f)
    \nonumber \\
    &=& \frac{7}{4} + \frac{1}{4} P(f) + \mbox{Re}\left(3 \varrho(1,f)-\varrho(1,D2)-\frac{3}{2} \varrho(D2,f) \right)
    \nonumber \\
    &=& \frac{7}{4} + \frac{1}{4}\mbox{Re}\left(3\varrho(1,f) + \varrho(2,f) - 5 \varrho(3,f) - 4 \varrho(1,P2)\right).
\end{eqnarray}
An inequality violation is observed whenever the KD terms used in the reconstruction of the state satisfy
\begin{equation}
\label{eq:criterion}
   3\varrho(1,f) + \varrho(2,f) - 5 \varrho(3,f) - 4 \varrho(1,P2) > 1. 
\end{equation}
Despite its negative KD terms, $\ket{T_{1f}}$ does not satisfy this requirement and therefore fails to violate the inequality in Eq.(\ref{eq:ineqality}). This is not surprising, since the probability of $S1$ in $\ket{T_{1f}}$ is zero, and the remaining probabilities in the sum $\Sigma$ can be divided into pairs from the same context. In general, the inequality given by Eq.(\ref{eq:ineqality}) cannot be violated when one of the five probabilities is zero. As pointed out in \cite{Kly08}, a state that optimally achieves a violation can be obtained by superpositions of all five states representing the outcomes that contribute to the inequality, simultaneously achieving very similar probabilities for all five outcomes. A state that is close to the maximal inequality violation is given by \cite{Hof23b}
\begin{equation}
    \ket{N_x} = \frac{1}{3}\left(2 \ket{1}+2 \ket{2} +\ket{3}\right).
\end{equation}
The tomographic data for this state is given by
\begin{eqnarray}
\label{eq:Nx}
    \varrho(1,f|N_x)&=& \hspace{0.2cm} 2/9, \hspace{1.2cm} \varrho(1,S2|N_x)=1/3,
    \nonumber \\
    \varrho(2,f|N_x)&=& \hspace{0.2cm} 2/9, \hspace{1.2cm} \varrho(1,P2|N_x)=-1/9,
    \nonumber \\
    \varrho(3,f|N_x)&=&-1/9.
\end{eqnarray}
The value of the sum $\Sigma$ obtained with this data is $\Sigma=20/9$, violating the inequality by $2/9$. The individual probabilities are $P(1)=P(2)=4/9$, $P(S1)=P(S2)=1/2$ and $P(f)=1/3$. 

Eq.(\ref{eq:criterion}) indicates that the violation of the non-contextual inequality in Eq.(\ref{eq:ineqality}) can be explained by the negative $KD$ terms of the paths $[3,f]$ and $[1,P2]$. The tomographic data of Eq.(\ref{eq:Nx}) thus explains the inequality violation in terms of its negative KD terms. However, Eq.(\ref{eq:ineqality}) does not require negative KD terms for its violation, and the specific selection of the tomographic data hides the fact that all five contexts play an equivalent role in the violation of the inequality. It is therefore desirable to derive a more complete form of the inequality, where the role of negativity and its relation to generalized probabilities as discussed in \cite{Spe08,Sha21} is more apparent. In the case of the $\ket{N_x}$ state, we would expect negative KD terms for all paths with four segments in table \ref{tab1}. In addition to the paths $[3,f]$ and $[1,P2]$, these are the paths $[2.P1]$, $[S1,D2]$ and $[S2,D1]$. If we simply complete the Kirkwood-Dirac distribution, we get the terms $\varrho(3,S2)$, $\varrho(2,P2)$ and $\varrho(3,P2)$. According to \ref{tab1}, each of these combinations of outcomes occurs in two of the eleven paths. The Hilbert space algebra confirms that the KD terms can be expressed in terms of the corresponding sums,
\begin{eqnarray}
\label{eq:separate}
\varrho(3,S2) &=& \varrho(S1,S2)+\varrho(D1,S2),
\nonumber \\ 
\varrho(2,P2) &=& \varrho(2,S1)+\varrho(2,P1),
\nonumber \\
\varrho(3,P2) &=& \varrho(S1,D2)+\varrho(0).
\end{eqnarray}
All of the KD terms defined by pairs of outcomes can be determined using Eq.(\ref{eq:KDterm}). The KD term $\varrho(0)$ of the path through all five unique outcomes needs a different definition. The general definition can be obtained by referring to the normalization of the Kirkwood-Dirac distribution,
\begin{equation} \label{eq:eleven}
\begin{split}
    \varrho(0)+ \varrho(S1,D2) + \varrho(D1,S2) + \varrho(2,P1) + \varrho(1,P2) + \varrho(3,f) & \\
    + \varrho(1,f)+ \varrho(2,f) + \varrho(1,S2) + \varrho(2,S1) + \varrho(S1,S2) & = 1.
\end{split}
\end{equation}
This equation can be simplified by summarizing the KD terms associated with $1$ and $2$, for
\begin{equation}
    \varrho(0)+ \varrho(S1,D2) + \varrho(D1,S2) + \varrho(3,f) + \varrho(S1,S2) = P(3).
\end{equation}
Independently, the fact that $\proj{S1}+\proj{D1}$ projects states into the Hilbert space spanned by $\ket{2}$ and $\ket{3}$, while $\proj{S2}+\proj{D2}$ does the same for $\ket{1}$ and $\ket{3}$ can be used to derive a very similar expression for $P(3)$,
\begin{equation}
P(3)= \varrho(S1,S2)+\varrho(D1,S2)+\varrho(S1,D2)+\varrho(D1,D2).
\end{equation}
Based on these relations, a conveniently symmetric definition of $\varrho(0)$ is
\begin{equation}
\label{eq:rhozero}
    \varrho(0) = \varrho(D1,D2) - \varrho(3,f).
\end{equation}
In general, $\varrho(0)$ contributes to all KD terms defined by pairs of outcomes that are unique to two different contexts. 

Eq.(\ref{eq:eleven}) shows that all quantum states can be characterized by a quasi probability distribution given by eleven KD terms that correspond to the 11 possible paths through the five contexts shown in table \ref{tab1}. If all KD terms are positive, these eleven terms can be interpreted as a probability distribution of the possible joint assignment of measurement outcomes to all five contexts. The eleven KD terms can thus establish a link between quantum statistics and hidden variable theories. However, quantum theory introduces additional relations between the eleven KD terms. Since only five KD terms are needed to characterize the quantum state completely, there should be a set of six relations that determine the remaining six KD terms as a function of these five. One of these relations is the normalization, and this relation can be used to determine $\varrho(0)$ once the other five KD terms have been determined. The remaining five relations can be derived from Eq.(\ref{eq:Lambdas}), where each KD term is given by the expectation value of the operator $\hat{\Lambda}(a,b)$ according to
\begin{equation}
   \varrho(a,b) = |\braket{a}{b}|^2 \mbox{Tr}(\hat{\rho}\hat{\Lambda(b,a)}).
\end{equation}
Eq.(\ref{eq:Lambdas}) thus defines a set of deterministic relations between the KD terms that express a distribution of statistical weights over the possible quasi realities obtained by assigning a specific measurement outcome to each context. These fundamental constraints satisfied by the KD terms of all possible input states are given by
\begin{eqnarray}
\label{eq:determinism}
2 \varrho(1,S2) + 3 \varrho(f,2) + 4 \varrho(D2,S1) &=& 1
\nonumber \\ 
2 \varrho(S1,2) + 2 \varrho(S2,1) + 3\; \varrho(3,f) \;\; &=& 1
\nonumber \\
3 \varrho(f,1) + 2\varrho(2,S1) + 4\varrho(D1,S2) &=& 1
\nonumber \\
4 \varrho(S2,S1) + 3\varrho(1,f) + 6\varrho(P1,2) &=& 1
\nonumber \\
3 \varrho(2,f) + 4\varrho(S1,S2) + 6\varrho(P2,1) &=& 1
\end{eqnarray}
The KD terms in each line correspond to paths that do not share the same measurement outcome in any of the five contexts. It is therefore possible to assign exactly one KD term to each outcome in each of the five lines. For outcomes that are shared by two contexts, this results in three non-zero KD terms, e.g. $\varrho(1,S2)=1/2$, $\varrho(1,f)=1/3$ and $\varrho(1,P2)=1/6$ for $\ket{1}$. Since the sum is one, the eleventh KD term is $\varrho(0)=0$. On the other hand, outcomes that are unique to only one context have a total of five non-zero KD terms. For the state $\ket{3}$, these are the four positive KD terms $\varrho(D2,S1)=1/4$, $\varrho(3,f)=1/3$, $\varrho(D1,S2)=1/4$, $\varrho(S1,S2)=1/4$, and the fifth negative KD term $\varrho(0)=-1/12$. 

The appearance of a negative KD term in states associated with a specific measurement outcome may seem surprising, since such a state is related to the other contexts directly by its inner products with the outcomes of those contexts. Indeed, $\varrho(0)$ is somewhat special, since it cannot be derived from a single weak value and does not appear in any Kirkwood-Dirac distribution of only two measurement contexts. It can be confirmed that, for all outcomes that are unique to only one context, the value of this KD term is $\varrho(0)=-1/12$. The origin of this value is found in a Bargmann invariant defined by the Hilbert space vectors representing the five outcomes that are unique to each context,
\begin{equation}
    \braket{3}{D1}\braket{D1}{P1}\braket{P1}{P2}\braket{P2}{D2}\braket{D2}{3} = - \frac{1}{12}.
\end{equation}
In principle, it is possible to show that Eq.(\ref{eq:rhozero}) identifies this Bargmann invariant with $\varrho(0)$ for the states $\ket{3}$, $\ket{D1}$, $\ket{P1}$, $\ket{P2}$ and $\ket{D2}$. However, it might be more useful to show that the value of $\varrho(0)$ must be the same for each of these states. This can be done by determining $\varrho(0)$ for a maximally mixed state, $\hat{\rho}=\hat{\rm I}/3$. For that state, $\varrho(a,b)=|\braket{a}{b}|^2/3$ and $\varrho(0)=-1/36$. Specifically, $\varrho(0)$ is negative because $|\braket{3}{f}|^2 > |\braket{D1}{D2}|^2$. Since $\varrho(0)$ is zero for each state shared by two contexts, the negative value of the maximally mixed state must correspond to one third of the value of the state unique to that context for each decomposition of the mixed state into a mixture of states representing the measurement outcomes of any of the three contexts. 

In general, all quantum states can be described by the $11$ quasi probabilities of the different paths through the contexts. However, even mixed states tend to have negative KD values. The only states that result in positive non-contextual probabilities are states that can be expressed as mixtures of the five outcomes that are shared by two contexts. Significantly, this does not include the maximally mixed state $\hat{\rm I}/3$ with its negative KD term of $\varrho(0)=-1/36$. The violation of non-contextuality as given by Eq.(\ref{eq:ineqality}) requires a very specific combination of negative KD terms. To find these, we can make use of the correspondence between KD terms and classical paths through the contexts. Path $[0]$ has no shared outcomes, paths $[1,P2]$, $[S1,D2]$, $[f,3]$, $[S2,D1]$, $[2,P1]$ have one shared outcome each, and the remaining five paths have two shared outcomes. The average number of shared outcomes can then be found by subtracting the KD terms of the paths with less than two shared outcomes from an initial value of two,
\begin{equation}
\label{eq:sigma}
    \Sigma = 2 - \varrho(S1,D2) - \varrho(D1,S2) - \varrho(2,P1) - \varrho(1,P2) - \varrho(3,f) - 2 \varrho(0).  
\end{equation}
This expression for the probability sum $\Sigma$ shows that a violation of inequality (\ref{eq:ineqality}) is impossible unless some of the KD terms given are negative. As suggested by Spekkens in \cite{Spe08}, the origin of contextuality paradoxes can be traced to negative KD terms that suppress the contributions of paths with one shared outcome or none at all. Importantly, Kochen-Spekker scenarios require a joint probability characterizing more than two contexts. The eleven KD terms introduced here achieve this for all five contexts, even though each KD term is defined by only two outcomes. This is the significance of table \ref{tab1}. Quantum contextuality actually provides an extended probability theory that corresponds to the assignment of outcomes to all five contexts at the same time.

\section{Contextual fluctuations}

The eleven KD terms represent a distribution of joint statistical weights for all five contexts, where each KD term represents a specific combination of outcomes for each of the five contexts, corresponding to one of the paths in table \ref{tab1}. Like other quasi probabilities, this multi-context Kirkwood-Dirac distribution seeks to bridge the gap between a naive realism that assumes a co-existence of the measurement outcomes along a path and an equally naive anti-realism that simply assumes that there is no connection between the outcomes of different measurements. To guard ourselves against both misconceptions, we should accept that the 11 element Kirkwood-Dirac distribution describes a deterministic relation between the five different measurement contexts that cannot be reconciled with any joint reality of outcomes. To understand the concept of quantum contextuality better, we need to focus on the fact that only one measurement context describes the reality of an actual event. Measurement probabilities are obtained from sums of KD terms for this event, and all of these sums will necessarily be positive. Negative KD terms can never be observed in isolation, since their experimentally observable effects appear only in sums where the contributions from positive KD terms outweigh those from the negative ones. Quantum contextuality suggests that the reality observed in one measurement context is fundamentally different from the reality observed in a different measurement context. However, the KD terms $\varrho(a,b)$ still describe a well-defined relation between the outcomes $a$ and $b$. We should consider this in more detail - if the relation between $a$ and $b$ is valid when we measure $a$, then the measurement of $a$ determines the reality of $b$ as well. The set of values provided by the KD terms thus describes a contextual reality of $b$ conditioned by the directly observed outcome $a$.

The problem that prevents us from understanding quantum mechanics is that we identify physical properties with their eigenvalue, clinging to the believe that these eigenvalues are a complete and exclusive set of possibilities from which we can construct an objective reality. However, this is a very unlikely interpretation of the formalism. Instead, we should consider a much wider variety of possibilities, where the outcomes of the context $a$ are related to contextual realities of outcomes $b$ that are different from the limited options of the binary yes or no associated with the eigenvalues of the projectors \cite{Hof21,Lem22,Hof23c,Mat23}. We need to accept that the values of projectors associated with other measurement outcomes can be different from $0$ or $1$. Is it possible to make intuitive sense of such a possibility? Perhaps a change of terminology is needed - ``truth value'' applies only to actual observations. Here, I will use the term ``outcome value,'' or more specifically, ``contextual outcome value''. It is easiest to visualize this change by considering the case of a single particle propagating along paths. As has already been demonstrated experimentally \cite{Lem22}, a particle is genuinely de-localized between different paths when these paths interfere in the output. We can now generalize this result to all quantum systems by describing the conditional reality of outcomes $b$ as a physical distribution of the total outcome value of $1$ over the outcomes $b$.  

The concept of a context dependent reality can be illustrated using the relations between measurement contexts described by the KD terms. Each measurement context has three outcomes, two of which are shared with other contexts and one of which is unique to that context. The measurement probabilities are given by sums of three KD terms for each of the shared outcomes, and by a sum of the remaining five KD terms for the outcome unique to this context. For instance, the probabilities of the context $\{\ket{1}, \ket{2}, \ket{3}\}$ are given by
\begin{eqnarray}
    P(1) &=& \varrho(1,f)+\varrho(1,S2)+\varrho(1.P2),
    \nonumber \\
    P(2) &=& \varrho(2,f)+\varrho(2,S1)+\varrho(2.P1),
    \nonumber \\
    P(3) &=& \varrho(3,f)+\varrho(S1,D2)+\varrho(D1,S2)+\varrho(S1,S2)+\varrho(0).
\end{eqnarray}
Each measurement outcome assigns a truth value of $1$ to one of the three outcomes, and a truth value of $0$ to the others. These truth values correspond to eigenvalues of the projection operators $\proj{a}$ of the measurement context. Quantum mechanics strongly suggests that we should not assign such eigenvalues to unobserved measurement outcomes. Instead, weak values can be used to relate the outcomes $b$ to a post-selected outcome of the measurement context $a$ \cite{Hof23,Hof21,Lem22}. Using the terminology above, these weak values are the contextual outcome values of $b$ in the context of $a$. If the measurement outcome $a$ is obtained with a probability of $P(a)$, the contextual outcome value $W(b|a)$ of an outcome $b$ is given by 
\begin{equation}
    W(b|a)=\frac{\varrho(b,a)}{P(a)}.
\end{equation}
Technically, $W(b|a)$  corresponds to the complex conditional probabilities discussed in some of my previous work \cite{Hof11,Hof12b}, where the status of the Kirkwood-Dirac distribution as a symmetric quasi probability of two measurement contexts was emphasized. However, contextuality suggests that we need to abandon the idea of a truth value of zero or one for outcomes that are not observed in their own context - or ``eigencontext'' as I shall refer to in the following.

If the contextual outcome values $W(b|a)$ are considered as individual values of the observable represented by the projector $\proj{b}$, it is possible to define the fluctuations of this value in the context of $a$ by assigning the probabilities $P(a)$ to the corresponding outcome values $W(b|a)$. The dependence of $W(b)$ on $a$ thus represents the fluctuations of $b$ in the context $a$. Since the average value of $\proj{b}$ is equal to the probability $P(b)$, the fluctuation of $W(b|a)$ is given by
\begin{equation*}
    \Delta W_b^2 = \sum_a (W(b|a)-P(b))^2 P(a).
\end{equation*}
As shown in a number of previous works, the fluctuations of weak values over a complete set of post-selected outcomes are equal to the uncertainty of the corresponding observable for pure state inputs \cite{Hof21,Shi10,Hal16}. For the contextual values $W(b|a)$,
\begin{eqnarray}
    \sum_a (W(b|a))^2 P(a) &\leq& P(b),
    \nonumber \\
    \Delta W_b^2 &\leq& P(b)(1-P(b)).
\end{eqnarray}
In the limit of pure states and precise measurements, the contextual fluctuations of the outcome values are equal to the fluctuations of the truth value observed in a direct measurement of the corresponding context. This is strong evidence that the values $W(b|a)$ describe a context dependent reality. 

To connect this analysis with quantum tomography, it is necessary to consider the role of the initial state in the definition of contextual fluctuations. If the state is a pure state, it is characterized by a deterministic statement corresponding to a certain prediction of a measurement outcome. Whether this prediction is explicit of not, it has an effect on the outcome values. An example may help to explain this dependence. Let us consider the state $\ket{P2}$. Its eleven KD terms are given by
\begin{align}
    \varrho(S1,S2|P2)&=0&    \varrho(3,f|P2)&=0 &
\nonumber \\
     \varrho(1,S2|P2) &= 0&  \varrho(2,P1|P2) &= 1/6& 
\nonumber \\
     \varrho(2,S1|P2) &= 1/2&  \varrho(1,P2|P2) &= 1/6& 
\nonumber \\
     \varrho(1,f|P2) &= 0& \varrho(S1,D2|P2) &= 1/4& 
\nonumber \\
     \varrho(2,f|P2) &= 0&  \varrho(D1,S2|P2) &= 0& \varrho(0|P2)=-1/12.&
\end{align}
As explained above, the KD term $\varrho(0)$ of this state is negative. We can therefore expect to find negative outcome values for some of the outcomes when the measurement context is not the eigencontext of that outcome. For instance, the values $W(D1|P2)$ of the outcome $\ket{D1}$ and their probabilities in the context $\{\ket{1}, \ket{2}, \ket{3}\}$ are given by
\begin{align}
    P(1|P2)&=1/6&   W(D1|P2;1)&=0&
    \nonumber \\
    P(2|P2)&=2/3&   W(D1|P2;2)&=1/4&
    \nonumber \\
    P(3|P2)&=1/6&  W(D1|P2;3)&=-1/2.&
\end{align}
Note that I have included the initial state in the conditions that determine the value $W(D1)$. This is intended as a reminder of the physical meaning of these values. The combination of state preparation $P2$ and measurement outcome $m$ completely describes the physical reality of the system. The value $W(D1)$ for this physical reality is determined by these conditions, leaving no uncertainty in the value of the projector $\proj{D1}$. We can confirm this assumption by  finding the fluctuation of $W(D1)$. In every measurement context, the average value of $W(D1)$ in the state $\ket{P2}$ is equal to the probability of $D1$, $P(D1|P2)=1/12$. The fluctuations of $W(D1)$ in $\ket{P2}$ are given by
\begin{equation}
    \Delta W_{D1}^2 = \left(\frac{1}{12}\right)^2 \frac{1}{6} + \left(\frac{1}{6}\right)^2 \frac{2}{3} + \left(\frac{7}{12}\right)^2 \frac{1}{6} = \frac{1}{12}\left(1-\frac{1}{12}\right).
\end{equation}
The fluctuations obtained for the context $\{\ket{1}, \ket{2}, \ket{3}\}$ are the same as the fluctuations obtained in the eigencontext of $\ket{D1}$, even though the outcome values themselves are very different. This is a strong indication that the individual realities associated with this fluctuation depend on the measurement context. In its eigencontext, the outcome $D1$ simply has a low probability. In the context $\ket{1}, \ket{2}, \ket{3}\}$, the main contribution to the expectation value of $P(D1)=1/12$ originates from the most likely outcome of $2$, for which the outcome value of $D1$ is $1/4$, much smaller than a truth value of $1$, but still three times larger than the average. The negative outcome value of $-1/2$ for the less likely outcome of $3$ is needed to achieve the same average and the same fluctuations as the eigencontext of $D1$. This is similar to the negative presence of a particle in a two-path interferometer that was reported in the neutron interference experiment in \cite{Lem22}. In general, negative contextual outcome values are observed when the probability of the observed outcome is low because of destructive interference between outcomes of the unobserved context. The quantum formalism assigns precise conditional values to these unobserved outcomes, providing a context dependent description of the quantum fluctuations of $D1$ in the input state $\ket{P2}$. Weak values and KD terms represent deterministic relations that allow statements about the values of unobserved physical properties. In the case of projectors, negative values indicate a particularly strong deterministic suppression of contributions from this outcome.  

The initial state defines averages and fluctuations. It does not define outcome values and probabilities. These can only be determined when the context is fixed. It is therefore interesting to take a look at states that seek to minimize the uncertainties if outcomes in all five contexts. As discussed above, the state $\ket{T_{1f}}$ achieves this by enhancing the KD term that corresponds to the path $[1,f]$ through the five contexts. Table \ref{tab3} already shows five of the eleven KD terms for this state. The complete set of eleven KD terms can be obtained from the table by making use of the observation that the $\ket{T_{1f}}$ state has $P(S1)=0$. Therefore, each of the eight elements in the table appear as one of the eleven KD terms, while the remaing three KD terms have values of zero. The complete set of KD terms is given by
\begin{align}
    \varrho(S1,S2|T_{1f})&= 0&    \varrho(3,f|T_{1f})&= 5/33&
\nonumber \\
     \varrho(1,S2|T_{1f}) &= 3/11&  \varrho(2,P1|T_{1f}) &= -2/33& 
\nonumber \\
     \varrho(2,S1|T_{1f}) &= 0&  \varrho(1,P2|T_{1f}) &=1/11 & 
\nonumber \\
     \varrho(1,f|T_{1f}) &= 5/11& \varrho(S1,D2|T_{1f}) &= 0& 
\nonumber \\
     \varrho(2,f|T_{1f}) &= 5/33&  \varrho(D1,S2|T_{1f}) &= -1/11& \varrho(0|T_{1f})=1/33.&
\end{align}
The state $\ket{T_{1f}}$ satisfies the conditions given by Eq. (\ref{eq:determinism}) by compensating the enhancement of $\varrho(1,f)$ above its eigenstate value of $1/3$ with two negative KD terms, $\varrho(2,P1)$ and $\varrho(D1,S2)$. Note that the paths $[2,P1]$ and $[S2,D1]$ represented by these two KD terms do not share any outcome with $[1,f]$ in table \ref{tab1}. This means that simultaneous control of five contexts is achieved at least in part by suppressing joint contributions of the unintended outcomes. We expect to see this effect more clearly in the relation between observed outcomes and contextual outcome values. 

We can now consider the statistics of outcome values for $f$, $S2$ and $P2$ in the measurement context $\{\ket{1}, \ket{2}, \ket{3}\}$. The probabilities of each outcome and the contextual distribution of the total outcome value of $1$ over $f$, $S2$ and $P2$ are given by
\begin{align}
    P(1|T_{1f})&=9/11&  W(f|T_{1f};1)&=5/9& W(S2|T_{f1};1)&=1/3& W(P2|T_{f1};1)&=1/9&
    \nonumber \\
    P(2|T_{1f})&=1/11&  W(f|T_{1f};2)&=5/3& W(S2|T_{1f};2)&=0& W(P2|T_{1f};2)&=-2/3&
    \nonumber \\
    P(3|T_{1f})&=1/11& W(f|T_{1f};3)&=5/3& W(S2|T_{1f};3)&=-1& W(P2|T_{1f};3)&=1/3.&
\end{align}
The first thing that should be noted is the way in which the high average of $P(f)=25/33$ for the outcome $f$ is obtained. Instead of assigning a high value to the most likely outcome of $1$, the Hilbert space formalism determines a value of only $5/9$ ($11/15$ of the average of $25/33$) to this outcome. Oppositely, the less likely outcomes $2$ and $3$ have a contextual value of $5/3$ for $f$, exceeding the eigenvalue limit of $1$ of the projector. This is a direct consequence of the relations in Eq.(\ref{eq:determinism}), where the high value of $P(f)$ is achieved by suppressing joint contributions of $(2,3)$ and $(S2,P2)$. This suppression is directly visible in the negative outcome values of $W(P2)$ for outcome $2$, and for $W(S2)$ for outcome $3$. If the distributions of outcome values are viewed together, the concentration of outcome values on $f$ can be observed for each of the outcomes, but it is much weaker for the most likely outcome of $1$ and extremely enhanced by outcome values outside the eigenvalue range for the less likely outcomes of $2$ and $3$. 

As noted above, the statistical fluctuations of the outcome values are all equal to the corresponding truth value fluctuations of the eigencontext statistics given by the probabilities of $P(f)=25/33$, $P(S2)=2/11$ and $P(P2)=2/33$. This can be confirmed by considering the differences between the outcome values and the average values represented by the probabilities,
\begin{eqnarray}
    \Delta W_{f}^2 = \left(\frac{20}{99}\right)^2 \frac{9}{11} + \left(\frac{10}{11}\right)^2 \frac{1}{11} + \left(\frac{10}{11}\right)^2 \frac{1}{11} = \frac{25}{33}\left(1-\frac{25}{33}\right),
    \nonumber \\
\Delta W_{S2}^2 = \left(\frac{5}{33}\right)^2 \frac{9}{11} + \left(\frac{2}{11}\right)^2 \frac{1}{11} + \left(\frac{13}{11}\right)^2 \frac{1}{11} = \frac{2}{11}\left(1-\frac{2}{11}\right),
    \nonumber \\
    \Delta W_{P2}^2 = \left(\frac{5}{99}\right)^2 \frac{9}{11} + \left(\frac{8}{11}\right)^2 \frac{1}{11} + \left(\frac{3}{11}\right)^2 \frac{1}{11} = \frac{2}{33}\left(1-\frac{2}{33}\right).
\end{eqnarray}
The statistical fluctuations of the outocomes $f$, $S2$ and $P2$ observed in their eigencontext are fully reproduced by the statistics of the outcome values associated with the outcomes $1$, $2$ and $3$. No additional randomness is required - the fluctuation of the outcomes in the context $\{\ket{1},\ket{2},\ket{3}\}$ has the same physical origin as the fluctuation of the outcomes in the context $\{\ket{f},\ket{S2},\ket{P2}\}$.  

As shown above, the inequality violations usually associated with quantum contextuality is a natural consequence of the dependence of outcome values on the measurement context, where negative contextual outcome values appear as fluctuations associated with low probability outcomes. Inequality violations can be achieved when a specific set of KD terms is negative, as shown in Eq.(\ref{eq:sigma}). In the case of the state $\ket{N_x}$, five of the six $KD$ terms contributing to the inequality violation are negative. The complete set of 
eleven KD terms can be obtained from the five terms in Eq.(\ref{eq:Nx}) and the relations in Eq.(\ref{eq:determinism}). The results read
\begin{align}
    \varrho(S1,S2|N_x)&= 1/4&    \varrho(3,f|N_x)&= -1/9&
\nonumber \\
     \varrho(1,S2|N_x) &= 1/3&  \varrho(2,P1|N_x) &= -1/9& 
\nonumber \\
     \varrho(2,S1|N_x) &= 1/3&  \varrho(1,P2|N_x) &=-1/9 & 
\nonumber \\
     \varrho(1,f|N_x) &= 2/9& \varrho(S1,D2|N_x) &= -1/12& 
\nonumber \\
     \varrho(2,f|N_x) &= 2/9&  \varrho(D1,S2|N_x) &= -1/12& \varrho(0|N_x)=5/36.&
\end{align}
The value of the sum $\Sigma$ is given by
\begin{equation}
    \Sigma(N_x)=2+3\times\frac{1}{9}+2\times\frac{1}{12}-2\times\frac{5}{36} = 2+\frac{2}{9}.
\end{equation}
The inequality (\ref{eq:ineqality}) is violated because the negative KD terms for paths with one shared outcome outweigh twice the positive KD term of the path with no shared outcome. The $\ket{N_x}$ state is nearly symmetric in the five contexts, with maximal suppression of the outcomes unique to each context. For the measurement context $\{\ket{1},\ket{2},\ket{3}\}$, the measurement probabilities and the contextual values $W(f)$, $W(S2)$ and $W(P2)$ are given by
\begin{align}
\label{eq:Nstat}
 P(1|N_x)&=4/9&  W(f|N_x;1)&=1/2&   W(S2|N_x;1)&=3/4&   W(P2|N_x;1)&=-1/4&
    \nonumber \\
 P(2|N_x)&=4/9&  W(f|N_x;2)&=1/2&  W(S2|N_x;2)&=0&   W(P2|N_x;2)&=1/2&
    \nonumber \\
 P(3|N_x)&=1/9&   W(f|N_x;3)&=-1&  W(S2|N_x;3)&=3/2&  W(P2|N_x;3)&=1/2&
\end{align}
The first thing to note regarding the dependence of the contextual values on the measurement outcomes is the dominance of suppression effects over enghancement effects. For an outcome of $1$, the contextual values of $f$ and $S2$ are nearly equal, but the contextual value of $P2$ is negative. For an outcome of $3$, the value of $S2$ is greater than one, but the value of $P2$ is also much larger than its average of $P(P2)=1/12$. More impressive is the suppression of the value of $f$ to $-1$, especially since the average value of $f$ is $P(f)=1/3$. For the outcome $2$, the value of $S2$ is the eigenvalue of zero, but it is worth noting that the conditional values of $f$ and $P2$ are the same. Each outcome has high values for two of the three outcomes, even though the statistics of these outcome values fully explain the input state uncertainties, leaving no room for any uncertainties in the conditional values themselves. Negative contextual values describe the deterministic relations between different contexts as extreme anti-correlations, exceeding the limits based on the assumption that the maximal correlation is a joint reality represented by a conditional value of one for a specific outcome and a value of zero for the two others.

It is important to remember that a non-contextual reality can only assign truth values of zero or one to each outcome, whether the outcome is actually observed in the measurement or not. The operator formalism contradicts this assumption on a seemingly technical level, by assigning weak values defined by the combination of deterministic input state properties with the additional information provided by the measurement context. The statistical analysis of the distribution over weak values for a specific measurement shows that weak values represent the same uncertainties also described by the eigenvalues, even though the individual values are different. In the case of the outcome $f$, the average value defined by the initial condition $\ket{N_x}$ is $P(f)=1/3$ in every context. In the eigencontext, a value of 1 is obtained with a probability of $1/3$ and a value of zero is obtained else. In the context $\{\ket{1}, \ket{2}, \ket{3}\}$, a value of $1/2$ is obtained with a probability of $8/9$, and a value of $-1$ is obtained with a probability of $1/9$. The fluctuations of the two contexts are the same,
\begin{equation}
    \Delta W_{f}^2 = \left(\frac{1}{6}\right)^2 \frac{8}{9} + \left(\frac{4}{3}\right)^2 \frac{1}{9} = \frac{1}{3}\left(1-\frac{1}{3}\right).
\end{equation}
The negative contextual value associated with a rare outcome of $3$ describes the fluctuations in terms of sudden deviations from a value much closer to the average than the truth values of zero and one. Both the values and the distributions of quantum fluctuations depend on the measurement context, even though their total magnitude is completely detemined by the initial state.

The outcome $S2$ has an average value of $P(S2)=1/2$, which corresponds to a maximal uncertainty of $1/4$ in any context. Since the outcome $\ket{2}$ is orthogonal to $\ket{S2}$, an eigenvalue of zero is obtained with a probability of $4/9$. The remaining values are $3/4$ with a probability of $4/9$ and $3/2$ with a probability of $1/9$. The fluctuations of $W(S2)$ are given by
\begin{equation}
    \Delta W_{S2}^2 = \left(\frac{1}{2}\right)^2 \frac{4}{9} + \left(\frac{1}{4}\right)^2 \frac{4}{9} + \left(1\right)^2 \frac{1}{9} = \frac{1}{2}\left(1-\frac{1}{2}\right).
\end{equation}
Here too, an extremal value is obtained for the low probability outcome. This time the fluctuation is up towards an outcome value greater than one. 

The lowest average value is that of $P2$, with only $P(P2)=1/6$. Similar to the statistics of $f$, there are only two possible outcome values of $P2$ in the context $\{\ket{1}, \ket{2}, \ket{3}\}$. A value of $1/2$ is obtained with a probability of $5/9$ and a value of $-1/4$ is obtained with a probability of $4/9$. The fluctuations are 
\begin{equation}
    \Delta W_{P2}^2 = \left(\frac{1}{3}\right)^2 \frac{5}{9} + \left(\frac{5}{12}\right)^2 \frac{4}{9} = \frac{1}{6}\left(1-\frac{1}{6}\right).
\end{equation}
Here, the effect of the context on the fluctuations is opposite to that observed for $f$. Since the low probability result $3$ and the high probability result $2$ have the same value, the probabilities for the two possible values are very similar. Due to this absence of rare events, the fluctuations are almost equally distributed over the different outcomes. Even the negative value of $-1/4$ is closer to the average of $1/6$ than the eigenvalue of $1$ obtained in the eigencontext. 

The relations between different measurement contexts used for quantum state reconstructions cannot be interpreted by statistical arguments alone. Pure states represent deterministic conditions that can be used to identify contextual values for unobserved outcomes by combining each measurement outcome with the initial conditions represented by the quantum state. Weak values describe this deterministic relation between measurement contexts, explaining in detail how each measurement resolves the uncertainties of measurement outcomes from different contexts in a consistent manner. Inequality violations and quantum paradoxes can all be explained as a natural consequence of this well-defined dependence of physical properties on the specific measurement context.

\section{Discussion and outlook}

The relations between five contexts in a three dimensional Hilbert space can be derived using suitable non-orthogonal expansions of quantum states. As the analysis in this paper has shown, the structure of contextuality can be separated from the statistics of Kirkwood-Dirac distributions by identifying relations between different KD terms that limit the number of independent elements in the Kirkwood-Dirac distribution to five. At the same time, it is possible to relate the KD terms to a simultaneous representation of all five contexts, where a total of eleven KD terms represent the eleven classical paths through the five contexts. The importance of representing more than two contexts in a single statistical expression cannot be stressed enough. Quantum phenomena are difficult to understand because we usually look at only one context at a time. Correlations between two contexts often appear to be sufficiently ambiguous to insert a wide variety of interpretations into the gaps. Two contexts are not sufficient for a description of deterministic relations since there are no conditions that these relations must satisfy. The cyclic relation between the five contexts shown in figures \ref{fig1} and \ref{fig2} solves this problem. We can then take a look at the differences between the classical determinism of joint realities and the quantum mechanical relations expressed by operators and state vectors. 

Each of the eleven KD terms in the five context scenario contributes its probability to exactly one outcome in each context. The relation between two different contexts can be described by a conditional distribution, where a different distribution of context $b$ is associated with each outcome in context $a$. Negative contextual values ensure that the uncertainties of the initial quantum state can be explained by observable fluctuations in the values of $b$ for different outcomes of $a$. This result suggests that the contextual values $W(b|a)$ describe conditional realities of $b$ associated with the initial state and the outcomes $a$ of the measurement context. It is important to remember that eigenvalues cannot be observed unless a strong measurement is actually performed \cite{Mat23}. This means that there is no information about the outcomes $b$ available anywhere in the universe whenever the measurement of $a$ is conducted instead. KD terms do not represent joint realities, since the outcomes in the argument cannot be obtained jointly. Instead, the KD terms characterize the relation between different measurements and the deterministic relations between physical properties that determine them. 

The results presented above can be used to settle a number of important questions regarding the foundations of quantum mechanics. First, they address the physics of quantum states and their representations. For each measurement context, a quantum state provides the correct probability distribution. However, the quantum state also defines the context dependence of physical properties by implicitly introducing deterministic conditions. In the end, quantum states represent the past of a system in a context independent manner. This means that quantum states neither describe the physical reality of the system, nor an ensemble of possible physical realities. Quantum states without measurements are fundamentally incomplete.

Second, the results address causality and determinism. The relation between different measurement contexts is fully determined for any pure state input. These relations are faithfully represented by the weak values that define contextual outcome values for each measurement context. It may be interesting to consider what the origin of the fluctuations in a pure state are. Since pure states cannot be divided into different possibilities, the physics of state preparation must erase all external information that could distinguish between different possibilities. A close analysis of interaction dynamics suggests that this problem combines aspects of entanglement with measurement theory \cite{Mat23}. 

Most importantly, the careful analysis of relations between different measurement contexts and their observable statistics confronts us with the difficulty of distinguishing between description and fact. It is clearly problematic that we identify quantum state components with measurement outcomes, regardless of whether the measurement actually happens or not. The temptation of fantasizing an imagined reality into such expressions is probably greater than we realize. The idea that a mathematical representation is an image of the object can only be maintained if the representation stays close to the actual observations recorded in an experiment. The original formulation of quantum mechanics strays too far into the realm of formal abstraction to permit such a naive identification. Hopefully, the present analysis of the relation between different quantum phenomena will pave the way towards a better practical understanding of the theory.

\end{document}